# Implementation of MPPT Technique of Solar Module with Supervised Machine Learning

Ruhi Sharmin, Sayeed Shafayet Chowdhury, Farihal Abedin, and Kazi Mujibur Rahman

*Abstract*— **Automated calibration of a maximum power point tracking (MPPT) algorithm and its efficacious implementation for the photo-voltaic (PV) system is pivotal for harnessing maximum possible energy from solar power. However, most existing calibration methods of such an MPPT system are cumbersome and vary greatly with the environmental condition. Hence, an automated pipeline capable of performing suitable adjustments and accurate analysis for solar PV systems is highly desirable. To counter this issue, numerous algorithms have been proposed so far, such as perturb and observe (P&O) method, incremental conductance (IC) method, fractional open circuit voltage (FOCV) method, short circuit current method, fuzzy logic based algorithm etc. While these approaches perform pretty well and produce overall acceptable results, recent surge of data-driven machine learning (ML) approaches hold great promise in this research domain. Since, supervised ML methods are trained directly based on the data, no human-designed heuristics are involved, which makes these techniques highly accurate and robust. As a result, in this paper, we proposed a method using supervised ML in solar PV system for MPPT analysis. For this purpose, an overall schematic diagram of a PV system is designed and simulated to create a dataset in MATLAB/ Simulink. Thus, by analyzing the output characteristics of a solar cell, an improved MPPT algorithm on the basis of neural network (NN) method is put forward to track the maximum power point (MPP) of solar cell modules. To perform the task, Bayesian Regularization method was chosen as the training algorithm as it works best even for smaller data supporting the wide range of the train data set. The theoretical results show that the improved NN MPPT algorithm has higher efficiency compared with the Perturb and Observe method in the same environment, and the PV system can keep working at MPP without oscillation and probability of any kind of misjudgment. So it can not only reduce misjudgment, but also avoid power loss around the MPP. Moreover, we implemented the algorithm in a hardware set-up and verified the theoretical result comparing it with the empirical data.**

## I. INTRODUCTION

With the advent of modern technology, renewable energy is the most talked topic in the world now-a-days mostly because of global energy crisis. Consequently, the search for green energy has led us to the various forms of renewable energy, they derive directly from the sun, or from heat generated deep within the earth. Included in the definition is electricity and heat generated from solar, wind, ocean, hydro power, biomass, geothermal resources, and bio-fuels and hydrogen derived from renewable resources. Among them, solar energy is abundant in nature and does not pose any threat for environmental hazard and hence the safest option as a reliable power source [1]. Engineers developing solar inverters implement MPPT algorithms to maximize the power generated by PV solar systems [2]. The algorithms account for factors such as variable irradiance (sunlight) and temperature to ensure that the PV system generates maximum power at all times making MPPT a crucial factor [3-6]. There are several algorithms for MPPT analysis but each of them can be categorized based on the type of the control variable it uses: (a) Voltage; (b) current; or (c) duty cycle. The main advantage of these algorithms is their capability and flexibility to solve non- linear problems. Moreover, they can generate the optimal solution or multi-peaks MPPT for global maxima with acceptable efficiency. So, these methods show superior tracking performance over conventional algorithms [7] [8].

Discussing previous works on MPPT algorithms first leads to Perturb and Observe (P&O) method which is very simple and basic yet unacceptable in many cases because of lesser accuracy [9] [10]. For example, in a real-life set up, it is assumed that the system does oscillate around the MPP indicating that a continuous perturbation in one fixed direction will lead to an operating point which will be very far away from the actual MPP. This process continues until the increase in insolation is reduced or eliminated completely. Among other existing methods, the name of Incremental Conductance (IC) method comes immediately after P&O method [11]. But despite higher accuracy in comparison with P&O method, it is not so easy to implement. Both perturb and observe, and the incremental conductance, are ideal examples of "hill climbing" algorithms that can find the local maximum power point of the power curve for the operating condition of the solar PV array. However, Fractional Open Circuit Voltage method and Fractional Short Circuit Current method are also popular because of improved efficiency than P&O method [12]. Fuzzy Logic is a bit complicated to implement but gives satisfying results [13] [14]. Also, fuzzy logic can reduce the slower tracking speed and the oscillation noise around Maximum power point (MPP) which are the two main disadvantages of the P&O method. But in this era of Artificial Intelligence and the boom of Neural Network architecture, using machine learning gives the most accurate results within shortest possible time span. As a result, no other MPPT algorithm can beat the level of efficiency provided by a neural network model of solar MPPT controller. This paper definitely supports this claim.

In this paper, an NN model has been proposed to solve the

Ruhi Sharmin is with the Department of Biomedical Engineering, Purdue University, 206 S. Martin Jischke Drive, West Lafayette, IN 47907-2032, USA (e-mail: rsharmin@purdue.edu).

Sayeed Shafayet Chowdhury is with the Department of Electrical & Computer Engineering, Purdue University, 501 Northwestern Ave., West Lafayette, Indiana 47907-2035, USA (e-mail: chowdh23@purdue.edu)

Farihal Abedin and Prof. Dr. Kazi Mujibur Rahman is with the Department of Electrical Engineering, Bangladesh University of Engineering & Technology (BUET), Dhaka, Bangladesh (e-mail: farihalabedin@pg.eee.buet.ac.bd and kmr@eee.buet.ac.bd).

same age old problem that is MPP tracking [15] [16]. The main challenge in using NN models lies in training them properly and hence the novelty of this paper lies in tackling that challenge successfully.

The main contributions of this paper are:
1. The complete dataset was prepared in such a way that it eliminates data bias while supporting a wide range of temperature and irradiance. Extra care has been taken to make it free from overfitting.
2. To ensure higher accuracy than the existing models, supervised machine learning approach is taken for this robust MPPT model.

Some existing models uses the data required to generate the ANN network obtained from the principle of Perturbation and Observation (P&O) method. Consequently, it pertains the errors that is generated from the P&O method [16]. Error back propagation method is also used in some models in order to train neural network. But in our model, we took a basic approach and used the basic equations of solar cell circuit model and that is why, chances of in-built errors are taken care of very well. Also, hardware support has been added to add to reliability of the simulation results.

## II. BACKGROUND

### A. Temperature Dependence and Effect of Irradiance

While using any form of renewable energy, how it impacts nature and how nature has its impact on it are some things we must keep under consideration. Same goes with using solar energy as a power source. As the intensity of solar power depends directly on the irradiance and temperature, we must consider the environmental factors while extracting the maximum power from solar panels.

The duration and intensity of sunlight that strikes the surface of the PV panel directly control the numerical value of the output current [17][18]. This explains the reason of popularity of solar energy as a form of renewable energy in the countries which are mostly tropical. But unlike the output current, the output voltage does not go under such direct changes due to change in irradiance.

The one factor that mostly affects output voltage and power is temperature. The relation between temperature and output voltage is linear, so to increase output voltage, higher temperature is preferable. As extra power in PV modules is decreased in high temperature, the advantage of MPPT charge controllers also decreases. So, even at normal operating temperatures, the additional unused power of an MPPT charge controller compared to a PWM controller can be minimal.

### B. Schematic of the Base MPPT Model

The main focus of this paper is in developing the controller rather than designing the DC-DC Converter. The problem is of supervised machine learning domain and we used regression to predict the value of maximum power point current ($I_{mpp}$) to track the maximum power point for a given irradiation G and temperature T [19].

Figure 1 shows that G and T are inputs for the PV panel which generates panel voltage ($V_{PV}$) and drives panel current ($I_{pv}$) to a DC-DC converter and a controller. The MPPT controller compares the value of maximum power point current ($I_{mpp}$) with $I_{PV}$ and generates the PWM switching signal (D) for the DC-DC converter to deliver maximum power to the load. Since there are multiple input variables, this will be multivariate regression [20]. Coupling to the load in order to transfer the maximum power requires providing a higher voltage or higher current. Usually, a buck boost scheme or sometimes a buck converter is used with a voltage and a current sensor tied into a feedback loop using a controller to vary the switching times of the switching signal [21].

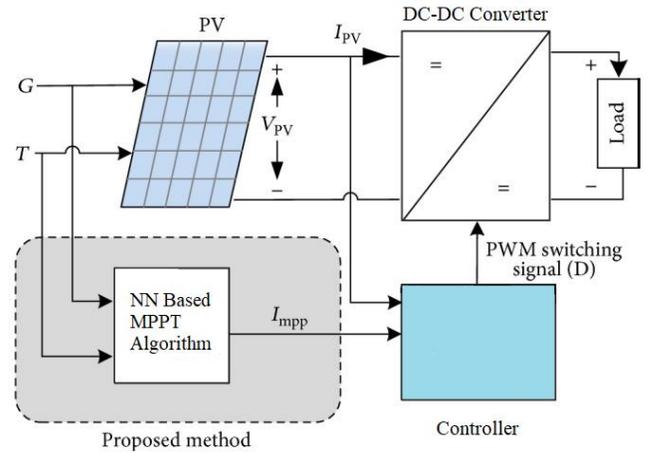

Figure1. Block diagram for the entire system

### C. Parameters of the Solar Module in Simulation

Our paper uses a simple and basic equivalent circuit model for the solar PV cell which consists of a real diode. The diode is placed in parallel with a current source but the current source is an ideal one. The ideal current source supplies current which is in proportion to the irradiance to which it is exposed. There are two conditions of interest for the actual PV and for its equivalent circuit. The first one is the current that flows when the terminals are shorted together (the short-circuit current, $I_{sc}$). The second one is the voltage across the terminals when the leads are left open (the open-circuit voltage, $V_{oc}$).

In most real cases, a more robust PV equivalent circuit model is needed where some resistive elements accounting for power losses, for example, a parallel leakage (or shunt) resistance $R_{sh}$ and a series resistance $R_s$ are included. Such a circuit is used in this paper and portrayed in Figure 2. In Figure 2, the electrical equivalent model for solar cell takes G and T as inputs and produce $I_{ph}$ which flows through three different branches. One portion of the current goes through a branch containing the diode, the other two portions flow through the two types of resistances [22]. The current produced from the solar PV panel, $I_{pv}$ is then supplied to the load.

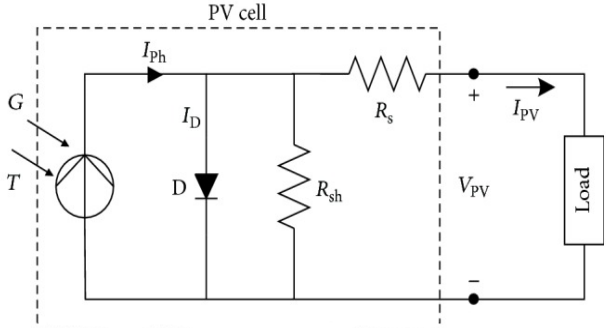

Figure 2. Electrical Equivalent Model of Solar Cell.

Here, $I_{ph}$ is photo current produced by solar power, $V_{pv}$ and $I_{pv}$ are the voltage and the current measured at the output terminal generated from solar PV panel respectively, $R_{sh}$ and $R_s$ are shunt and series resistances respectively, $I_D$ is the diode current and D is used to denote the diode.

Each solar panel, or module, is rated to produce a certain portion of wattage, a fixed amount of voltage and a definite level of amperage under specific conditions (such as-irradiance, air mass, cell temperature). But in ideal cases, all cells possess more or less similar characteristics with no mismatching losses [23]. The electrical characteristics of the PV module that are used in this paper along with temperature is shown in Table I.

TABLE I: PV MODULE CHARACTERISTICS

| parameter | value |
|---|---|
| n | 54 |
| $R_s$ $(\Omega)$ | 0.2210 |
| Rsh $(\Omega)$ | 415.4050 |
| Isc (A) | 8.21 |
| Voc (V) | 32.9 |
| T (º C) | 25 |

Here, n is used to denote the number of series-connected solar cells in the PV array, $R_s$ and $R_{sh}$ stand for series and shunt resistance respectively, $I_{sc}$ and $V_{oc}$ are used for short-circuit current and open-circuit voltage respectively and T is used to denote atmospheric temperature.

### D. Building Blocks of the Base Model

The first step of creating the base model for making database and mathematical modelling of solar PV array in MATLAB/

Simulink requires the use of different equations for obtaining the values of different electrical characteristics of the equivalent circuit model.
The required equations for the 5 sub-system blocks of the main Simulink block are stated below.

Photo Current:

$$I_{PH} = \frac{\{I_{SC} + K_I\ (T - 298)\}G}{1000} \qquad (1)$$

Here, $I_{PH}$: photo-current (A); $I_{SC}$: short circuit current (A); $K_I$: short-circuit current of cell at 25 °C or 298K and 1000 W/m$^2$; T: operating temperature (K); G: solar irradiation (W/m$^2$).

Saturation Current:

$$I_0 = (\frac{I_{RS} \times T^3}{T_N^3})\ e^{\{\frac{q\ E_{G0}\left(\frac{1}{T_N} - \frac{1}{T}\right)}{nK}\}} \qquad (2)$$

Here, $T_N$: nominal temperature = 298.15 K; $E_{G0}$: band gap energy of the semiconductor = 1.1 eV;
q: electron charge = $1.6 \times 10{-}19$C; n: the ideality factor of the diode; K: Boltzmann's constant, = $1.3805 \times 10{-}23$ J/K.

Reverse Saturation Current:

$$I_{RS} = \frac{I_{SC}}{e^{\{\frac{q \times V_{OC}}{n \times N_S \times K \times T}\}} - 1} \qquad (3)$$

Here, $V_{oc}$: open circuit voltage (V); $N_s$: number of cells connected in series.

Current through Shunt Resistance (Shunt Current):

$$I_{SH} = \frac{V + I \times R_S}{R_{SH}} \qquad (4)$$

Output PV Current:

$$I = I_{PH} - I_0 \times I_{SH} \times [e^{\{\frac{q \times (V + I_{RS})}{n \times K \times N_S \times T}\}} - 1] \qquad (5)$$

The next step is to concatenate the 5 sub-system blocks and build the main block to get output current and voltage from a solar module. Figure 3 shows that there are in total 5 separate sub system blocks for reverse saturation current (input T and output $I_{RS}$ or $I_{rs}$), saturation current (inputs T, $I_{RS}$ and output $I_0$), photo current (inputs T, G and output $I_{PH}$ or $I_{ph}$), shunt current (inputs I, V and $I_{SH}$ or $I_{sh}$) and PV current (inputs V, T, $I_0$, $I_{PH}$, $I_{SH}$ and output I).

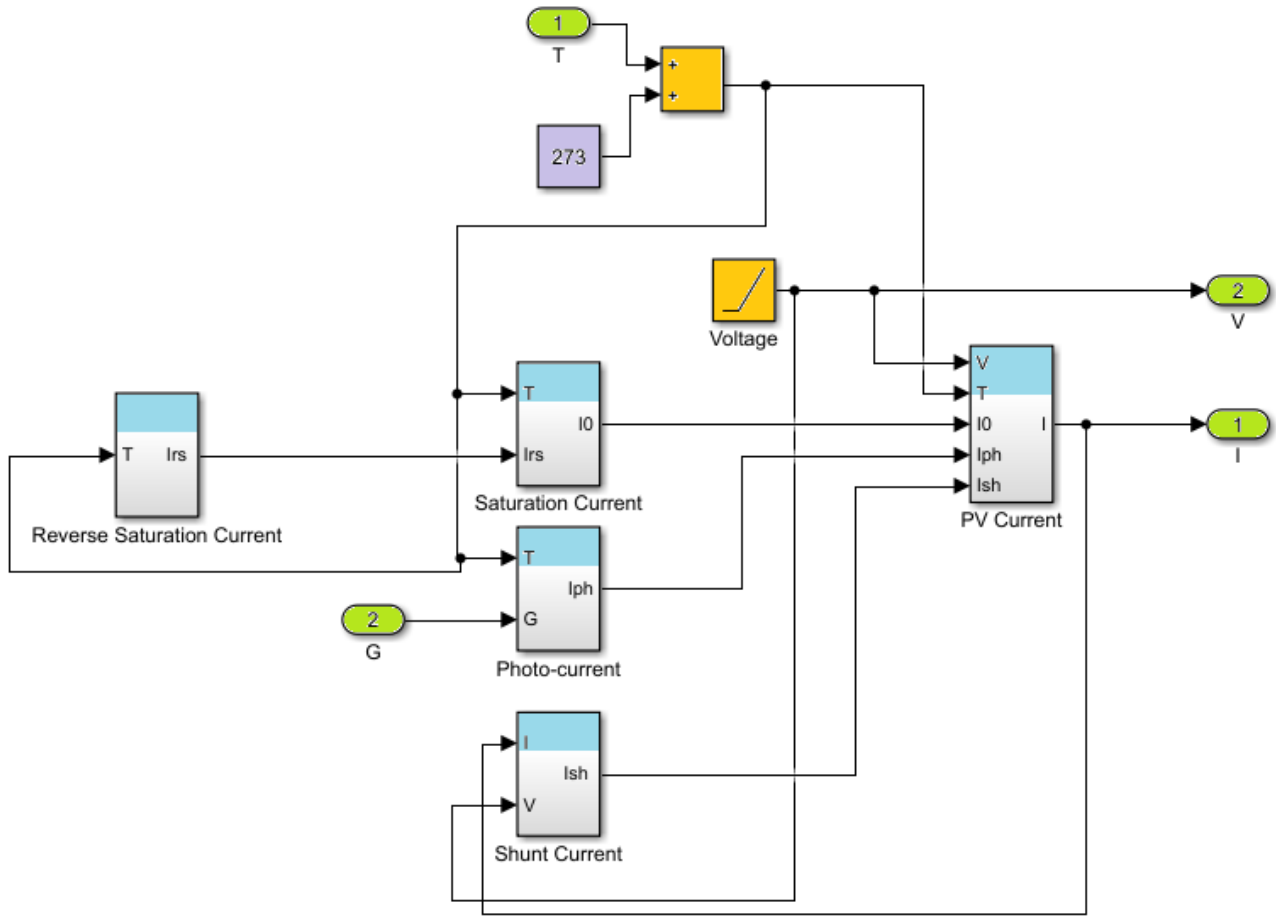

Figure 3: Main block connecting the 5 sub system blocks

## III.   DESIGN OF THE MPPT MODULE

### A.   *Proposed Methodology*

The design of the MPPT module can be divided into three steps as shown in Figure 4. In the first step, the basic PV module takes temperature and Irradiance values as input data and gives current value at MPP as output data [24][25]. The second step is data collection which relies on the simulation results from the first step. After the train dataset is created, it can be used to train the proposed multilayer perceptron (MLP) type artificial neural network (ANN) and get the final output from our test data [26].

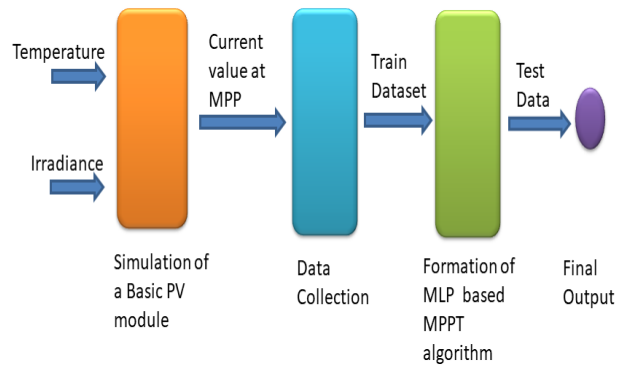

Figure 4: Workflow Diagram of the Proposed Methodology

## B. Simulation Results of the Base Model

The PV Array block of MATLAB Simscape library implements an array of photovoltaic (PV) modules. The array is built of strings of modules, these modules are connected in parallel, each string consisting of modules connected in series. This block allows us to model preset PV modules from the National Renewable Energy Laboratory (NREL) System Advisor Model (Jan. 2014) as well as PV modules that we define.

The PV Array block is a five parameter model using a current source $I_L$ (light-generated current), diode ($I_0$ and $n_I$ parameters), series resistance $R_s$, and shunt resistance $R_{sh}$ to represent the irradiance- and temperature-dependent I-V characteristics of the modules. Figure 5 shows that the solar PV module takes T and G as inputs, and then the module generates output current and voltage as denoted by I and V, respectively. The values of I and V are used to generate an I-V graph and P-V graph for better speculation of the MPP of the system.

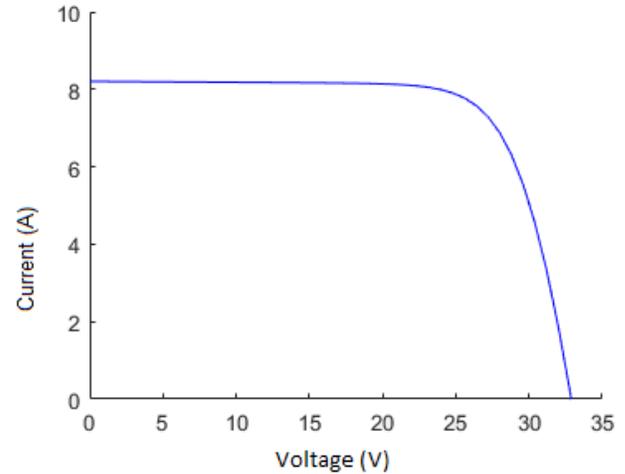

Figure 6: Current vs. Voltage graph

Figure 4 shows the power versus voltage graph which clearly denotes the MPP, the voltage value corresponding to the highest value of power denotes $V_{mpp}$.

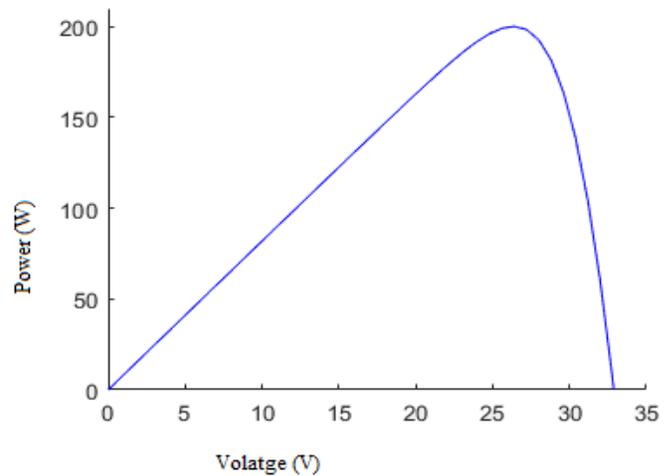

Figure 7: Power vs. Voltage graph

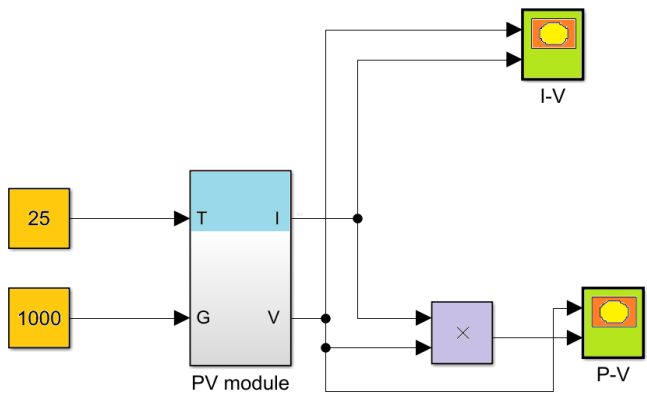

Figure 5: Schematic Diagram of the output

Let, $I_r$ = 1000 W/m$^2$ and T = 25 ºC

Now, after simulating the prescribed model, the graphs (I vs V and P vs V) derived from this simulation perfectly matches with the known shapes of PV model outputs.

This implies that this model works correctly and the data obtained from this model can be safely used for further investigation and for training a neural network architecture later on.

Figure 4 shows the current versus voltage graph which clearly denotes the MPP, the current value and the voltage value at the MPP can easily be perceived from this graph.

We know, any solar cell has $I_{sc}$ (here, 8.1A) and $V_{oc}$ (here, 32.9V) fixed. Here, the values of $V_{mp}$ at $P_{max}$ can manually be found for inputs: T=25K & G=1000 W/m2.

At power, Pmax=200.0170W (found by scrolling down all the power values, as this is the biggest data among all) at time index=2.6400, we get $V_{mp}$=26.4000 V at the corresponding time. This way, the values of T and G can be changed and corresponding values of $V_{mp}$ at $P_{max}$ is calculated. Finally, the value of current at maximum power point ($I_{mp}$) is obtained from this data or a short MATLAB code can be generated to do the task in a short span of time.

For this cell, every necessary data can be found from the workspace of MATLAB. At this stage, the value of $V_{mp}$ (voltage at maximum power) at $P_{max}$ is required. To collect this data, a simple workaround (Simulation Data Inspector>Log selected Signals) is performed.

### C. Creation of a Customized Dataset

After running total 1,300 simulations, we created our own dataset for T starting from 15 unit to 40 unit (40-15+1=26 values of T); (for each T, the values of G are 200,210,220….1090 etc. totaling in 50 values). In real atmospheric conditions, the difference between maximum and minimum temperature can be large for a certain geographical area which is taken care of by the large range of temperature values used in the dataset.

To make the data randomized, it is modified using simple mathematical techniques. This helps prevent data bias for the neural network model. The basic three operations used are shuffling, splitting and transposing.

Shuffle: All the collected data is shuffled randomly.

Splitting: In this paper, the data matrix size is 1300×3. Now, the data file is split into two different files:

    (i) The input file includes T & G data,

    (ii)The output file includes $I_{mp}$ data.

Transpose: As per the requirement of Matlab's neural fitting (nftool), the features (i.e. inputs) have to be in the columns of the data matrix. The input matrix must be transposed which includes T and G values. Similarly, the target matrix (i.e. output) must also be transposed. Finally, a column matrix which includes $I_{mp}$ values can be obtained in this process.

### D. Fitting Data into Neural Network Architecture:

In this paper, a multilayer feed-forward network was used as the neural network architecture. Here, each layer of nodes receives inputs data from the previous layers. The outputs of the nodes in one layer are inputs to the next layer. Moreover, the inputs to each node are combined using a weighted linear combination. Finally, the result is modified by a nonlinear function before being output.

As per definition, this neural network is formed in three layers, called the input layer (passive nodes), hidden layer (active nodes), and output layer (active nodes), the first layer is the input and the last layer is the output. Figure 6 shows that the neural network architecture used in this model consists of 15 hidden neurons in the hidden layer.

In this MLP NN, adaptive moment estimation (Adam) has been used as the optimization algorithm instead of the classical stochastic gradient descent procedure [27]. For the loss function, mean squared error (MSE) method has been applied. The number of epoch and the value of learning rate are 1000 and .001 second (approximately) respectively.

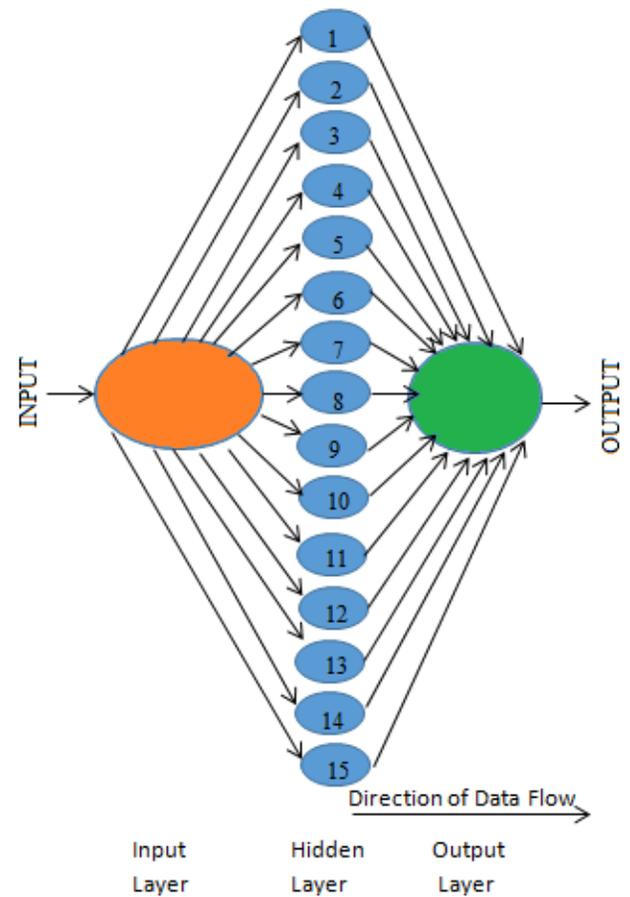

Figure 8: block diagram of NN architecture model.

At first, both the input (2 variables) and output (2 variables) data files was loaded. Then 5% of our data was selected as test dataset which includes 65 samples, 85% data as train dataset which includes 1105 samples, 10% data as validation dataset which includes 130 samples.

### E. Use of BRANN for Regression Analysis

Bayesian Neural Network, BNNs are important in specific settings, especially when the factor of uncertainty is taken care of. Some examples of these cases are decision making systems, (relatively) smaller data settings, Bayesian Optimization, model-based reinforcement learning and other similar areas.

In our model of MPPT, we used Bayesian regularized artificial neural networks (BRANNs) because of its higher level of robustness in comparison with standard back-propagation nets. Another reason for using BRANN is that it can lessen or eliminate completely the need for lengthy cross-validation. In fact, Bayesian regularization is a very sturdy mathematical tool that converts a nonlinear regression into a well-posed statistical problem statement in the manner of a ridge regression. So, Bayesian Regularization method was chosen as the training algorithm as it works best even for smaller data.

## IV. EXPERIMENTAL RESULTS

After creating and training the network, a very small mean square value of error was observed which ensures the effectiveness of the model. The MSE (mean square error) value for testing is very low (~2.87X10-3) for the proposed model.

### A. Simulink Deployment

Simulink deployment of the model has made it a very handy tool which instantly shows MPP after setting the input parameters. Moreover, the time lag to generate the output is very negligible hence establishing the model as a fast one.

Figure 9 shows that if we insert T and G values (Here, T=25º and $I_r$ = 1000 W/m$^2$) as inputs to the function fitting NN model, it instantly shows the corresponding current value at MPP (Here, I_mp(A) value = 7.592 A).

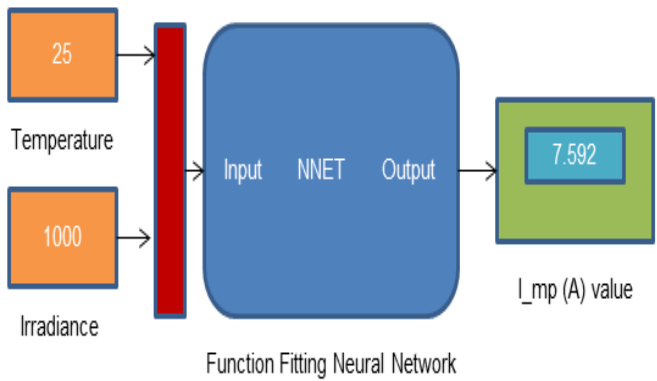

Figure 9. Simulink Deployment of NN Model.

### B. Bias Values & Weight Values in the NN Architecture

As an additional parameter, bias values in the NN architecture help adjust the output along with the weighted sum of the inputs to the neuron. Another important use of bias values is that it helps shift the activation function in either right or left direction. Bias values for input in hidden layer is portrayed in the following table. Figure 10 shows the NN architecture which takes input for the inner hidden layer, then comes the output layer and finally the output can be obtained at the last stage. W and B indicates the weight values and the bias values respectively.

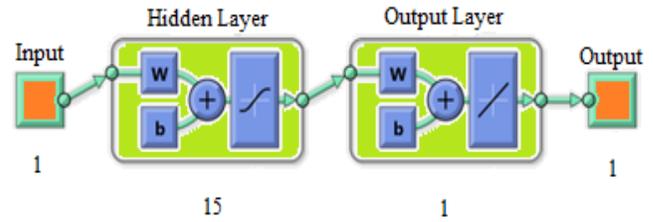

Figure 10. NN architecture with weight and bias values.

All the bias values and weight values can be readily obtained from the model and are mentioned in table II, III and IV which will make the hardware implementation of the proposed MPPT model possible in no time.

TABLE II. BIAS VALUES OF INPUT IN HIDDEN LAYER:

| Bias Values For Input In Hidden Layer |
| --- |
| -0.0788662358905827 |
| -0.199252461268044 |
| 0.476703785811994 |
| 0.0788437136945187 |
| 0.0788437137489292 |
| -0.0843897428619086 |
| 0.0788437137616096 |
| -0.0788437136323596 |
| 0.0788437137854139 |
| -0.0511645156192877 |
| 0.0521386721602554 |
| 0.0788437137594562 |
| 0.0788437137550028 |
| 0.325165283314198 |
| 0.199252475098178 |

Another important parameter in the NN architecture is weights, which represent the strength of the connection between units. If the weight value of node 3 is higher than that of node 4, the neuron 3 has higher influence over neuron 4. Hence, weight values can bring down the importance of the input values. Weights of input in hidden layer are portrayed in the following table.

TABLE III. WEIGHTS OF INPUT IN HIDDEN LAYER

| T | G |
| --- | --- |
| 0.330659943126136 | 0.375354867765757 |
| -0.243061055602675 | -0.302898742825237 |
| -0.413686729890811 | 0.124969405112560 |
| 0.0503908825102565 | -0.242889973645394 |
| 0.0503908825336595 | -0.242889973660344 |
| 0.286942170255656 | -0.871006037765083 |
| 0.0503908825390176 | -0.242889973663920 |
| -0.0503908824834618 | 0.242889973628136 |
| 0.0503908825490957 | -0.242889973670382 |
| -0.194105547886448 | -0.882406133751217 |
| -0.685142210229390 | 0.629560967118233 |

| 0.0503908825379756 | -0.242889973663218 |
|---|---|
| 0.0503908825362639 | -0.242889973662112 |
| 0.796782784260382 | -0.181097405343480 |
| 0.243060836205972 | 0.302898505989682 |

TABLE IV. WEIGHTS IN OUTPUT LAYER

| Weights In Output Layer |
|---|
| 0.526055556926590 |
| -0.526712907336645 |
| 0.560221719680131 |
| -0.357268781908311 |
| -0.357268781972241 |
| -0.335049187325717 |
| -0.357268781987367 |
| 0.357268781834757 |
| -0.357268782015078 |
| 0.371791504344763 |
| -0.318328545469931 |
| -0.357268781984634 |
| -0.357268781979911 |
| -0.244907095063019 |
| 0.526712952341475 |

| BIAS VALUE IN OUTPUT LAYER: | 0.1528 |
|---|---|

In an NN model, a neuron first computes the weighted sum of the inputs. For the inputs $(x_1, x_2, x_3,\ldots, x_n)$ and the weights $(w_1, w_2, w_3, \ldots, w_n)$, a weighted sum is computed as:
$x_1w_1 + x_2w_2 + x_3w_3 \ldots + x_nw_n$
Subsequently, a bias (constant) is added to the weighted sum:
$x_1w_1 + x_2w_2 + x_3w_3 \ldots, + x_nw_n +$ Bias
Finally, the computed value is fed into the activation function, which then prepares an output. Output, $Y = \Sigma$ (Weight × Input) + Bias. Activation function: $(x_1w_1 + x_2w_2 + x_3w_3 \ldots, + x_nw_n +$ Bias) These functions are mathematical tools that can normalize the inputs.

## C. Checking the Accuracy of the Result:

Bin represents the number of vertical bars observed on the graph. Figure 7 shows that the total error from neural network ranges from 0.2222 (leftmost bin) to 0.1968 (rightmost bin). This error range is divided into 20 smaller bins, so each bin has a width of (0.1968- (-0.2222))/20 = 0.02095.

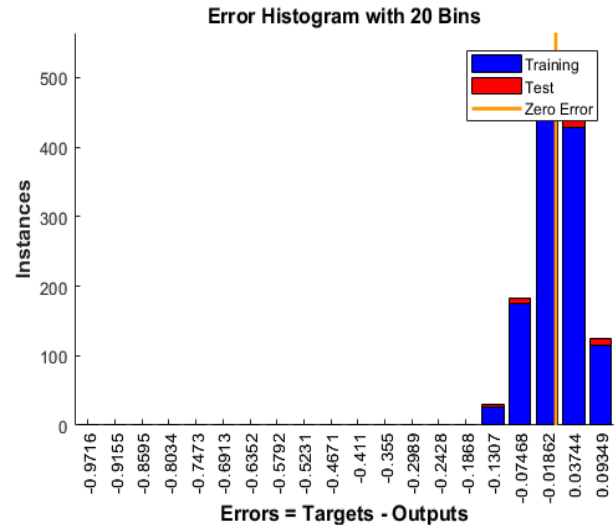

Figure 9. Error histogram of the output data.

Each vertical bar represents the number of samples from the dataset, which lies in a particular bin. For example, at the left half of the graph, there is a bin corresponding to the error of -0.07468 and the height of that bin for validation dataset is 200. It means that 200 samples from the dataset (Training+ Test) have an error which lies in the following range:
((-0.07468 - 0.02095/2), (-0.07468 + 0.02095/2))
The range (-0.012135, 0.008815) is even less than the range of the bin corresponding to -0.07468.

Figure 8 shows three plots for the regression analysis, the plot for training data shows that only one data point deviates from the line of regression, the plot for test data shows that almost all of the data points are either on very close to the line of regression and there is an overall third plot for the whole dataset.

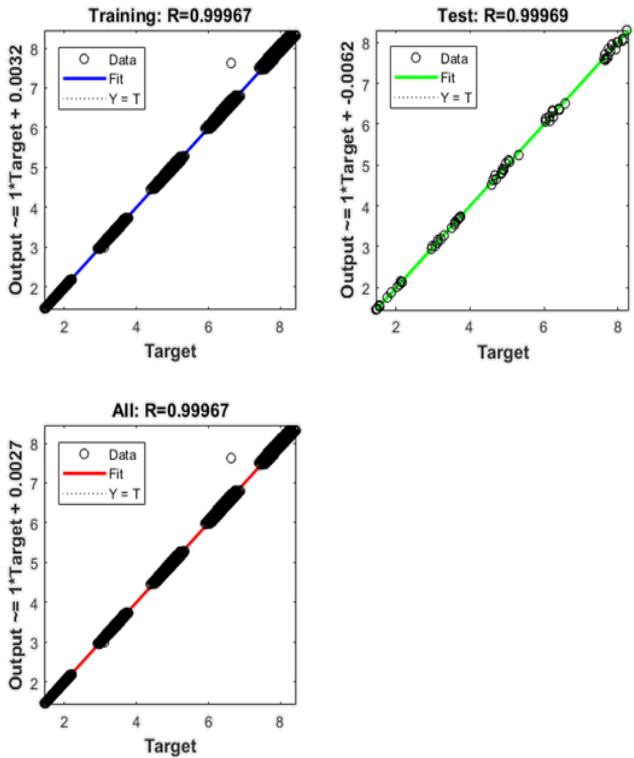

Figure 10. Plot for Regression Analysis.

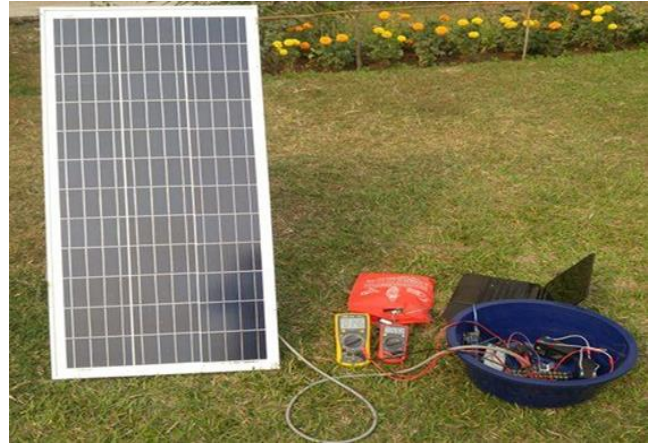

Figure 11: Full hardware set-up

The result of the regression analysis shows:
(i) Training: Only one data point deviated from the locus of the straight line
(ii) Test: No data point deviated from the locus of the straight line
(iii) Validation: Error was zero for validation dataset.

*Verification:*

Although this model is a software based one, it was implemented in a hardware setting to check the acceptability of the results [28] [29]. Figure shows the simple hardware set-up where one solar panel, one lead acid battery (100AH), two rheostat, two multimeters, cables, one solar panel controller were used. Also, a boost regulator was used here and the bias values and the weight values were used with the help of an Arduino board. The dimensions of the panel was1006 x 676 x 30 mm, the other panel specifications are as follows:

TABLE V. SOLAR PANEL SPECIFICATIONS

| Parameter | Value |
|---|---|
| Model Type | ZM-A-M-100 |
| Cell Type | 156mm x 104mm |
| Cell Arrangement | 36 cells in series |
| Dimensions(mm) | 1006 x 676 x 30 mm |
| Material | Monocrystalline Silicon |
| Maximum Power | 100W |
| Number of Cells | 36pcs |

*Analysis:*

Since, the proposed NN model is a simple one layer network, it can be implemented in real time. Also, temperature and irradiance values do not change rapidly in real environmental conditions [30]. Hence, the timing analysis shows that it takes only around 1 millisecond to produce the output and thus proving the model to be very time efficient.

In real time settings, noise gets added in every step of the MPPT analysis procedure, for example, while receiving input data through the sensors. Moreover, human errors also get involved during manual data collection. To counter this noise issue, some additive white Gaussian noise (AWGN) was added during the training phase and then finally trained the model with the data corresponding to the clean result and received better performance with test data. So, it shows less discrepancy even with very noisy data and thus adding to the robustness of the model.

For countries with extreme weather condition, the difference between the maximum temperature value in the summer and the minimum temperature value in the winter can be a large number. In such cases, the dataset used in this paper can easily be modified and some temperature values close to the maximum and minimum temperature values of that specific region of the world can be added. Similar task can be done for varying irradiance levels for remote places of the world. This easy but important modification of the dataset can add to the flexibility of the model.

The simpliest method for MPPT analysis so far has been the P&O method but the proposed model, being a very simple feed forward network, can easily replace.

The low-cost hardware set up proves that this model is very cost effective.

*Performance Comparison:*

Discrepancy: From the simulation data, for T=25K & G=1000W/m²: Imp=7.5764A but the neural network model produces Imp=7.592A. Percentage of deviation from simulation data= (7.592-7.5764)/ 7.5764X100% = 0.206%
So, this neural network model for tracking maximum power point is very effective. Increasing the number of neurons in hidden layer gives even better results (by adjusting network size). This model data was used to calculate the accuracy also.

Efficiency: Percentage of accuracy of the model data= (1-0.206%) = 99.794%. This high accuracy is one of the novelties of this paper. The size of the hidden layers to produce such a close approximation of the MPP is only 15.

For this same environmental condition, other popular and easy-to-implement methods give much lesser efficiency, for example, P&O has an accuracy level of 67.4%. Efficiency slightly increases in case of IC method (above 80%) but nowhere near compared to our NN model. If Fuzzy logic based algorithm is used, the efficiency can be increased up to as high as 96% despite the model becomes very complex to implement and time-lag becomes a major concern [31-33]. Usually, the NN models show efficiency of around 98%. But the proposed model shows higher efficiency (99.8% approximately) than even the other existing NN models of MPPT analysis for the same temperature and irradiance value.

## V. CONCLUSION & SCOPE OF FUTURE WORK

In this paper, an overall schematic diagram of a photovoltaic system is designed. By analyzing the output characteristic of a solar cell, an improved MPPT algorithm on the basis of NN method is put forward to track the MPP of solar cell modules. The theoretical results show that the improved NN MPPT algorithm has higher efficiency compared with the P&O method in the same environment, and the photovoltaic system can keep working at MPP without oscillation and misjudgment. So it can not only reduce misjudgment, but also avoid power loss around the MPP.

Partial shading can be included as the third input parameter to the proposed NN model.